\documentclass[,final]{aipproc}
\layoutstyle{6x9}
\usepackage{graphicx}
\newcommand{\abs}[1]{\left| #1\right|}
\newcommand{\fnd}[2]{\frac{\textstyle #1}{\textstyle #2}}
\newcommand{\fsml}[2]{\mbox{$\frac{#1}{#2}$}}
\newcommand{\bm}[1]{\mbox{\boldmath $#1$}}
\newcommand{\dissum}[2]{\displaystyle \sum_{#1}^{#2}}
\begin{document}
\title{Complex Meson Spectroscopy}
\author{Eef van Beveren}{
address={Centro de F\'{\i}sica Te\'{o}rica, Departamento de
F\'{\i}sica, Universidade, P3004-516 Coimbra, Portugal}
}
\author{Frieder Kleefeld}{
address={Centro de F\'{\i}sica das Interac\c{c}\~{o}es
Fundamentais, Instituto Superior T\'{e}cnico, Edif\'{\i}cio Ci\^{e}ncia,
P1049-001 Lisboa, Portugal}
}
\author{George Rupp}{
address={Centro de F\'{\i}sica das Interac\c{c}\~{o}es
Fundamentais, Instituto Superior T\'{e}cnico, Edif\'{\i}cio Ci\^{e}ncia,
P1049-001 Lisboa, Portugal}
}
\begin{abstract}
We do meson spectroscopy by studying the behavior of $S$-matrix poles in the
complex-energy plane, as a function of the coupling strength for $^3\!P_0$
quark-pair creation. Thereto, a general formula for non-exotic hadron-hadron
scattering involving arbitrary quark confinement is used, which can be applied
to all flavors. We find two distinct
types of poles, which we call {\it confinement} \/and {\it continuum} \/poles,
respectively.  Together, they suffice to understand the experimental meson
spectrum.
\end{abstract}

\maketitle

\begin{figure}[htbp]
\begin{tabular}{ccc}
\hline
\multicolumn{1}{|c|}{\includegraphics[height=100pt]{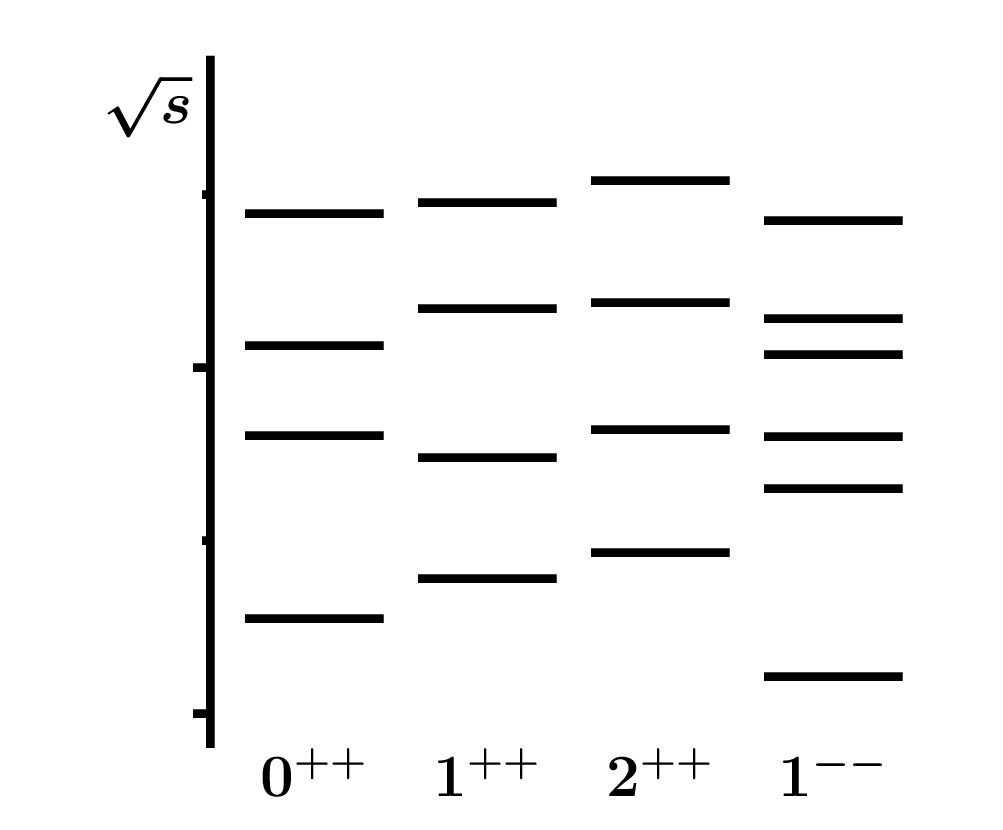}} &
\multicolumn{1}{c|}{\includegraphics[height=100pt]{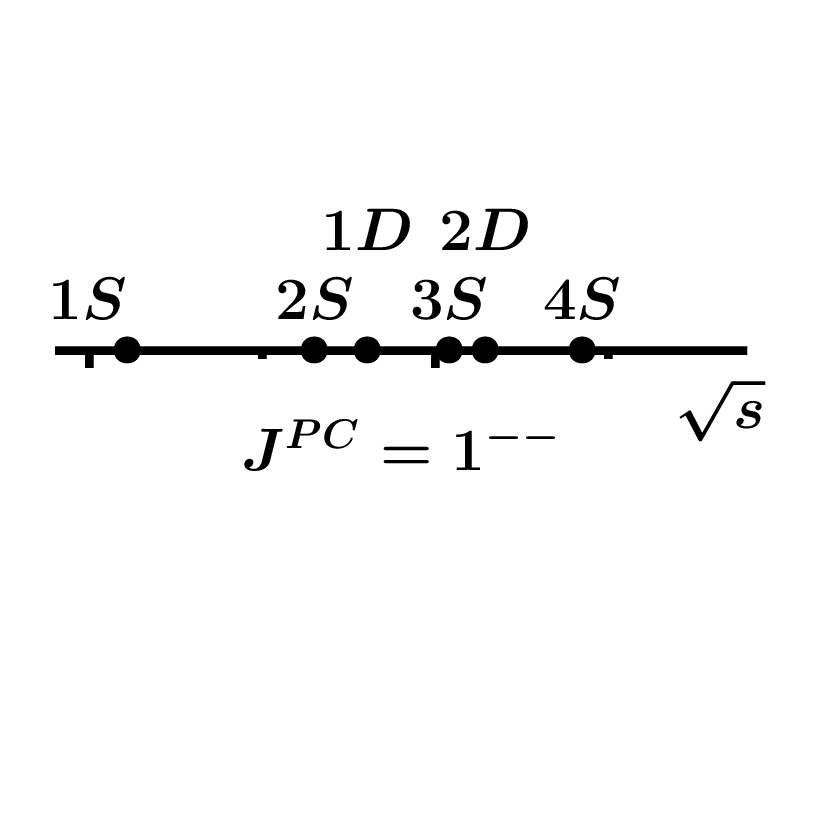}} &
\multicolumn{1}{c|}{\includegraphics[height=100pt]{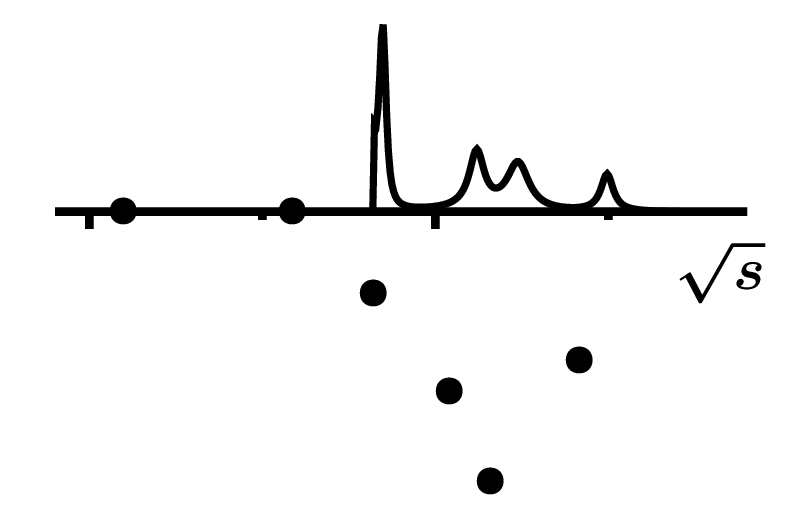}}\\
\hline
(a) & (b) & (c)\\
\end{tabular}
\caption{({\bf a}): Spectra as we are used to.
({\bf b}): The same, but restricted to one set of quantum numbers.
({\bf c}): A modern spectrum of
bound states below and resonances above threshold,
in terms of complex poles of the scattering amplitude
(Note that the units along the real and the imaginary axis
are not the same in this figure).}
\label{spectra}
\end{figure}

{\bf Introduction.}
From past studies on atomic spectra, we got so used to the terms
``line spectrum'' and ``states'' that we also apply them to mesonic spectra.
Such a situation is depicted in Fig.~\ref{spectra}a.
The decay widths are thought of as higher-order
effects, which indicate the life times of unstable states.
This is indeed a fruitful strategy for atomic spectra.
Starting from QED, one first determines the properties of the force
which binds electrons to the nucleus, and next
the corresponding bound-state spectra and states.
Subsequently, one determines life times of states
by considering the electromagnetic (EM) interaction of the photon field
with the electrons.

Seen from a different angle, the above strategy amounts to
separating the interaction of electrons and the nucleus with
the EM field into two distinct parts: the one binding
electrons and the nucleus into an atom, and the one
describing the scattering of photons off the atom.

Here, we apply the same technique to strong interactions,
but with the important difference that no manageable scheme exists
to disentangle the two parts when setting out from QCD.
Hence, they must be dealt with on the same footing.
As a consequence, also the central resonance position
and the resonance width must be described on the same footing,
which can be achieved by referring to the complex pole position of the
corresponding scattering amplitude.
Such situation is depicted in Fig.~\ref{spectra}c.

Furthermore, the term ``states'' is not very well defined for
resonances having large widths compared to their level separations.
One may, of course, refer to the state at a certain energy, say the
central resonance position.
But it is not reasonable to assume that at energies say 100 MeV higher
or lower, the state of the system is the same. Moreover, different
resonances may be overlapping.

Here, we shall be studying non-exotic meson-meson scattering.
Through quark-pair creation, this implies multi-channel scattering,
as well as intermediate states of confined systems of quarks and antiquarks.
Hence, the advantage of complex spectra is apparent,
since the pole structure of the full scattering matrix is the same
for all channels, although the structure of peaks and zeros in the various
channels can be very different.
\vspace{10pt}

{\bf Meson-meson scattering amplitude.}
On rather general grounds \cite{PRD27p1527,HEPPH0304105pure},
it can be shown that a system of confined quarks and antiquarks
coupled to free two-meson channels is well described
by a partial-wave ($\ell$) inverse cotangent matrix of the form
\begin{equation}
K_{\ell}(p)\; =\;
\fnd
{
\lambda^{2}\;
\dissum{n=0}{\infty}
\fnd{{\cal J}^{\ast}_{n\ell}(p)\;{\cal J}_{n\ell}(p)}{E(p)-E_{n\ell}}
}
{
\lambda^{2}\;
\dissum{n=0}{\infty}
\fnd{{\cal J}^{\ast}_{n\ell}(p)\;{\cal N}_{n\ell}(p)}{E(p)-E_{n\ell}}
-1
}
\;\;\; .
\label{Kmat}
\end{equation}
Here, $\lambda$ parametrizes the intensity of quark-pair creation,
$p$ stands for the relativistic relative on-shell momentum in one of the
two-meson channels, and $E(p)$ is the total invariant mass of the system.
Furthermore, $E_{n\ell}$ ($n=0$, 1, $\dots$) represent the {\it confinement
spectrum}, which is of course model-dependent, since it cannot be measured
directly, while ${\cal J}_{n\ell}$ and ${\cal N}_{n\ell}$ are convolution
matrix integrals over quark-antiquark distributions, meson-meson scattering
wave functions, and a transition potential.

If we approximate the latter potential by a radial delta function,
then we find for the $ij$-th matrix element of the partial-wave
scattering amplitude
\begin{equation}
\left[ T_{\ell}\right]_{ij}(p)\; =\;
\fnd
{
\lambda^{2}\;
\left\{
\dissum{n=0}{\infty}
\fnd{r_{i}(n)r_{j}(n)}{\sqrt{s}-E_{n\ell}}\;
\right\}\;
2a
\sqrt{\fnd{\mu_{i}\mu_{j}}{p_{i}p_{j}}}\;
p_{i}p_{j}
j_{\ell_{i}}\left( p_{i}a\right)j_{\ell_{j}}\left( p_{j}a\right)
}
{
1+\lambda^{2}
\dissum{m=1}{N}
\left\{
\dissum{n=0}{\infty}
\fnd{\abs{r_{m}(n)}^{2}}{\sqrt{s}-E_{n\ell}}\;
\right\}\;
2ia\mu_{m}p_{m}
j_{\ell_{m}}\left( p_{m}a\right)
h^{(1)}_{\ell_{m}}\left( p_{m}a\right)
}
\;\;\; .
\label{Tij}
\end{equation}
Here, $j_{\ell}$ represents the radial Bessel function,
$h^{(1)}_{\ell}$ the radial Hankel function of the first kind,
$a$ the delta-shell radius,
$\sqrt{s}$ the total invariant mass of the system, and
$\mu_{i}$ the reduced mass in the $i$-th two-meson channel.
The coefficients $r_{i}(n)$ ({\it vertex functions}),
which stem from the convolution integrals, are discussed later.
\vspace{10pt}

\begin{figure}[htbp]
\begin{tabular}{ccc}
\hline
\multicolumn{1}{|c|}{\includegraphics[height=120pt]{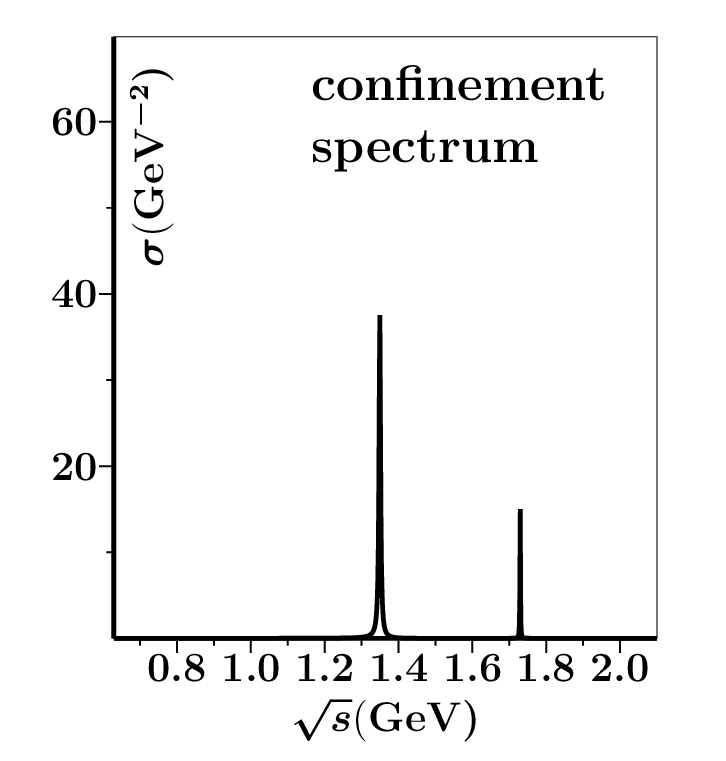}} &
\multicolumn{1}{|c|}{\includegraphics[height=120pt]{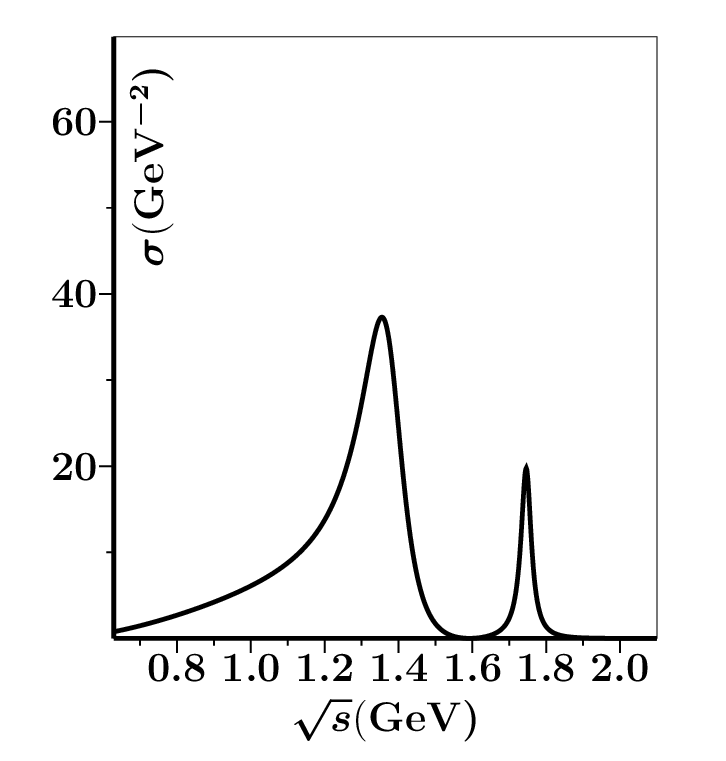}} &
\multicolumn{1}{|c|}{\includegraphics[height=120pt]{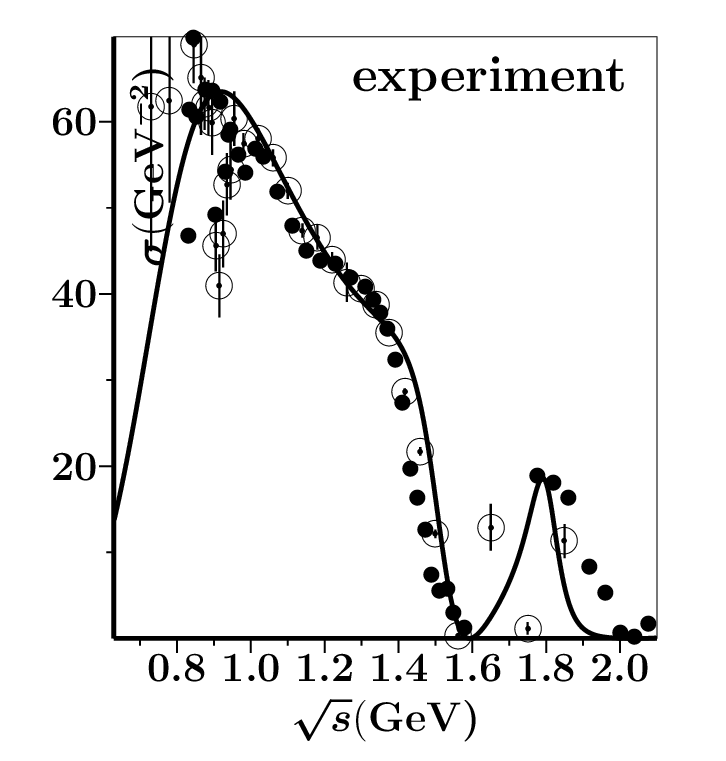}}\\
\hline
(a) & (b) & (c)\\
\end{tabular}
\caption{Cross section for $S$-wave isodoublet $K\pi$ scattering.
Left: For very small values of $\lambda$, one observes
the $J^{PC}=0^{++}$ $n\bar{s}$ confinement spectrum.
Middle: When $\lambda$ takes about half its model value, one notices
some more structure for low invariant masses.
Right: At the model's value of $\lambda$, this structure is dominant and
well in agreement with the experimental observations.
The data are taken from Ref.~\cite{NPB133p490} (open circles)
and Ref.~\cite{NPB296p493} (full circles).
}
\label{KpiS}
\end{figure}

{\bf Cross sections and poles.}
In Fig.~\ref{KpiS} we show how cross sections
following from formula (\ref{Tij}) vary with increasing $\lambda$,
for $S$-wave isodoublet $K\pi$ scattering.
In Fig.~\ref{KpiS}a the nonstrange-strange ($n\bar{s}$) confinement
spectrum is well visible for small $\lambda$,
whereas in Fig.~\ref{KpiS}c, for the model value of $\lambda$,
experiment is reproduced.

\begin{figure}[htbp]
\begin{tabular}{cc}
\hline
\multicolumn{1}{|c|}{\includegraphics[height=120pt]{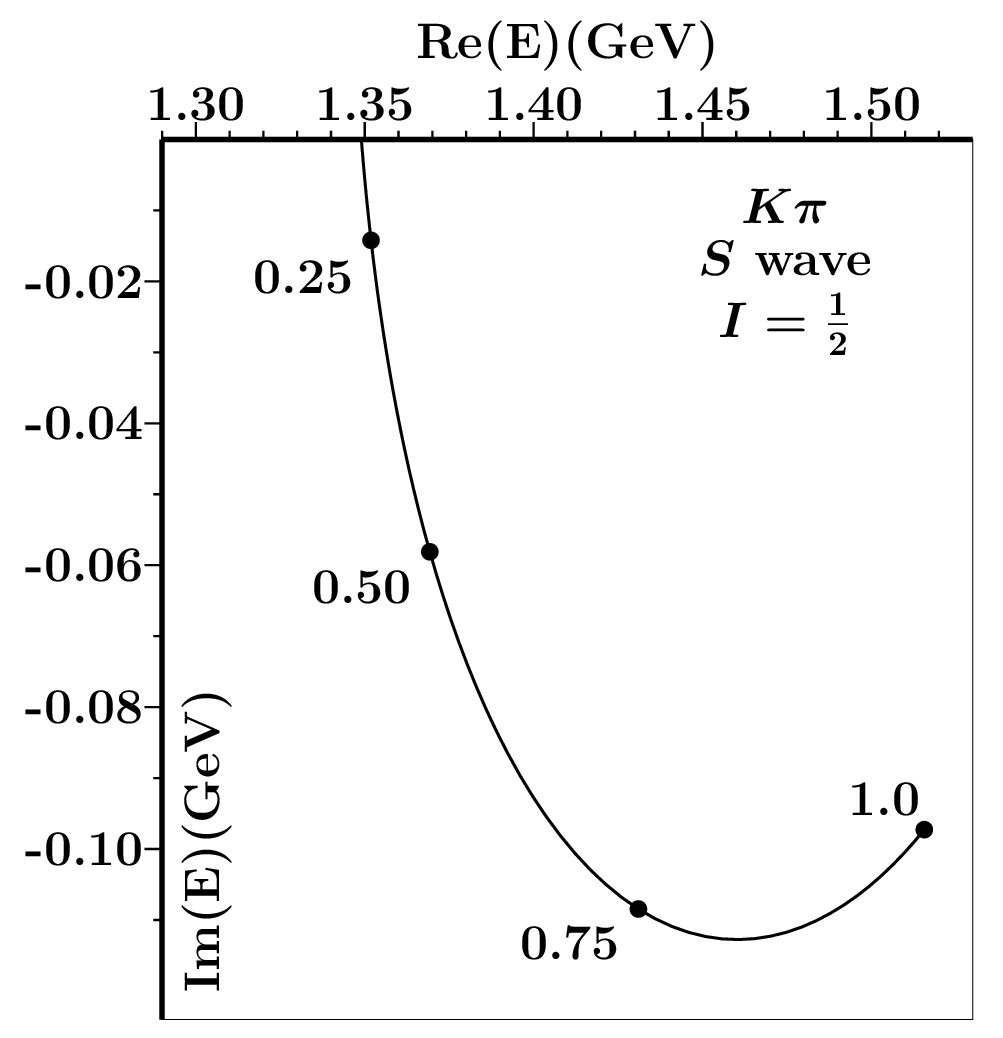}} &
\multicolumn{1}{|c|}{\includegraphics[height=120pt]{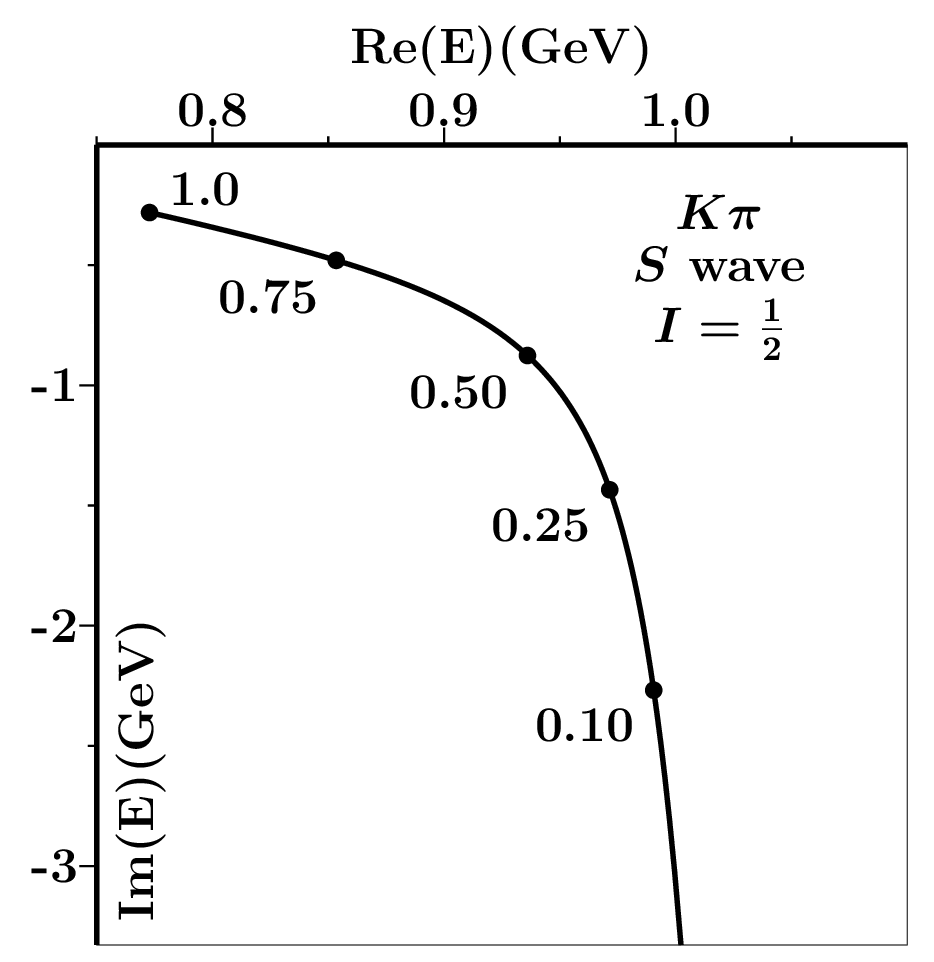}}\\
\hline
(a) & (b)\\
\end{tabular}
\caption{Pole positions for $S$-wave isodoublet $K\pi$ scattering,
as a function of the strength of quark-pair creation,
parametrized by $\lambda$ (see Eqs.~(\ref{Kmat},\ref{Tij})).
({\bf a}): Example of a {\it confinement pole},
which for $\lambda=0$ lies on the real axis at total invariant mass
$E_{0\ell}$, i.e., the ground state of the $n\bar{s}$
$J^{P}=0^{+}$ confinement spectrum.
({\bf b}): Example of a {\it continuum pole},
which for $\lambda=0$ has an infinite negative imaginary part,
i.e., the $K\pi$ continuum.
}
\label{KpiS12}
\end{figure}

The pole structure for varying $\lambda$ is depicted in Fig.~\ref{KpiS12}.
In Fig.~\ref{KpiS12}a we see how the pole stemming from the
ground state of the confinement spectrum moves through the complex-energy
($E=\sqrt{s}$) plane when $\lambda$ changes.
The pole movement shown in Fig.~\ref{KpiS12}b is very different.
For small $\lambda$ the pole disappears into the $K\pi$ continuum,
which is the reason why it is not visible in Fig.~\ref{KpiS}a.

\begin{figure}[htbp]
\begin{tabular}{cc}
\hline
\multicolumn{1}{|c|}{\includegraphics[height=120pt]{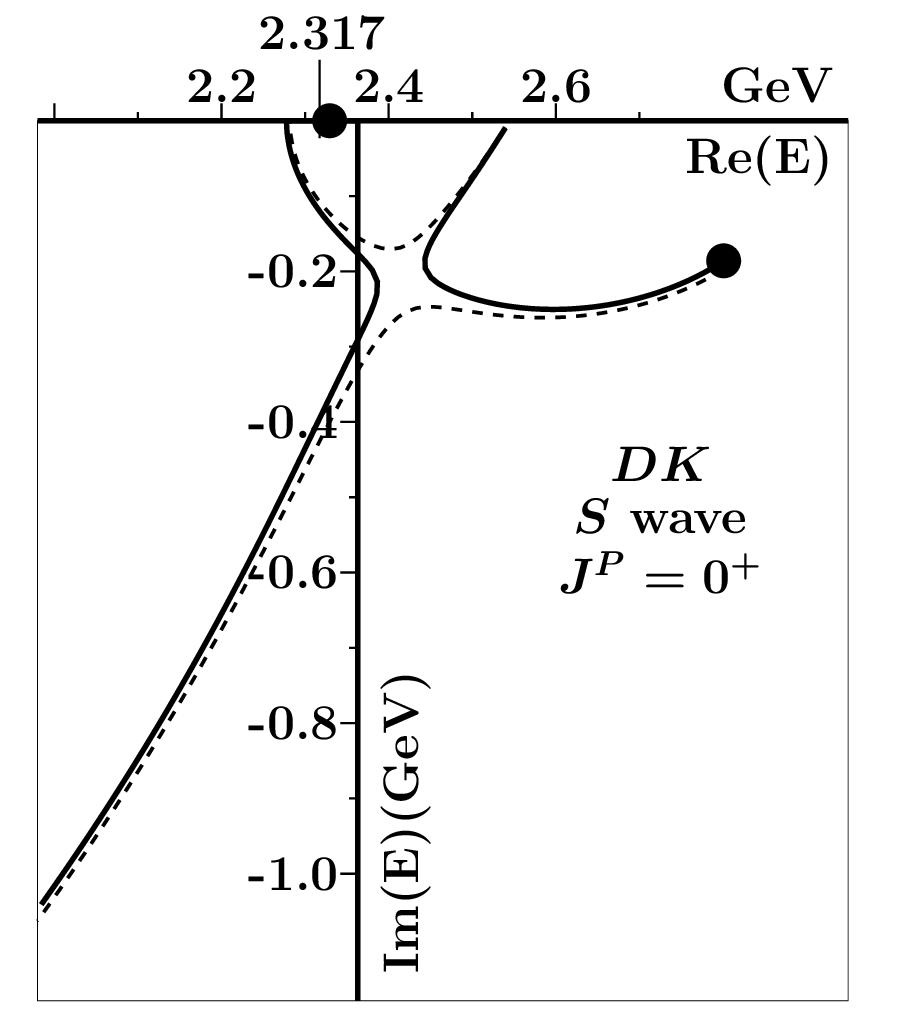}} &
\multicolumn{1}{|c|}{\includegraphics[height=120pt]{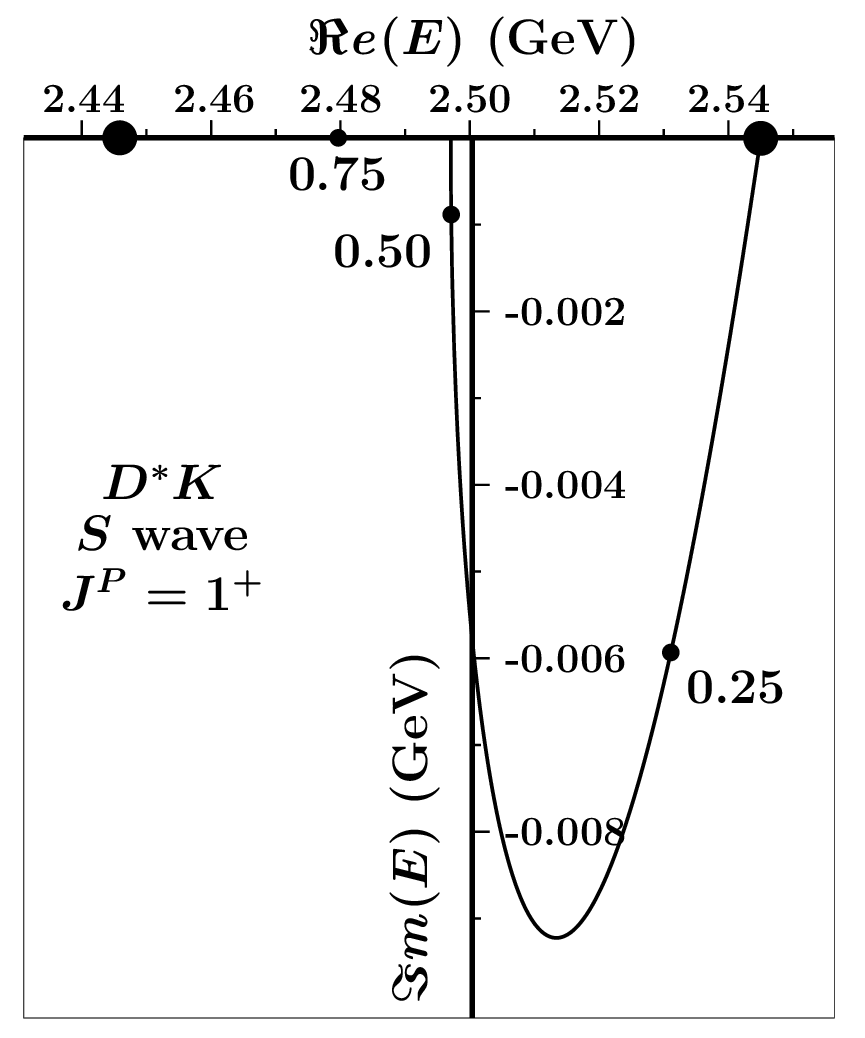}}\\
\hline
(a) & (b)\\
\end{tabular}
\caption{The lowest-lying poles for
$S$-wave $J^{P}=0^{+}$ $DK$ ({\bf a}) and $J^{P}=1^{+}$ $D^{\ast}K$ ({\bf b})
scattering.
The fat dots represent the pole positions for the model value of $\lambda$.
}
\label{KDSpoles}
\end{figure}

The lowest-lying scattering-amplitude poles
for $DK$ in $J^{P}=0^{+}$ and $D^{\ast}K$ in $J^{P}=1^{+}$
are shown in Fig.~\ref{KDSpoles}.
The two poles in Fig.~\ref{KDSpoles}a represent
the $D_{s0}^{\ast}(2317)$ resonance, which, in the absence of
isospin-violating processes, comes out on the real axis,
plus a broad resonance at about 2.8--2.9 GeV, undetected so far.

The two poles shown in Fig.~\ref{KDSpoles}b represent
the $D_{s1}(2460)$ and $D_{s1}(2536)$ resonances,
which both come out on the real axis.

In Fig.~\ref{KDSpoles}a we depict two possible alternatives
(one pair of solid lines and one pair of dashed lines).
They result from very small variations in the delta-shell radius.
Consequently, it is not at all trivial how to connect experimental
poles with either confinement or continuum poles.

The two poles in Fig.~\ref{KDSpoles}b both stem from
the ground state of the $J^{P}=1^{+}$ charm-strange confinement
spectrum, since there are two such states, i.e., $^{3\!}P_{1}$ and
$^{1\!}P_{1}$, which come out degenerate in our confinement spectrum.
The linear combination 33\% $^{3\!}P_{1}$ plus 67\% $^{1\!}P_{1}$
decouples from $D^{\ast}K$ $S$-wave scattering,
hence becomes a bound state in the scattering continuum
(see also Ref.~\cite{PRD72p054029}). On the other hand,
the orthogonal linear combination of $^{3\!}P_{1}$ and $^{1\!}P_{1}$
couples fully, and turns into a bound state below threshold.
\vspace{10pt}

\begin{figure}[htbp]
\includegraphics[height=129pt]{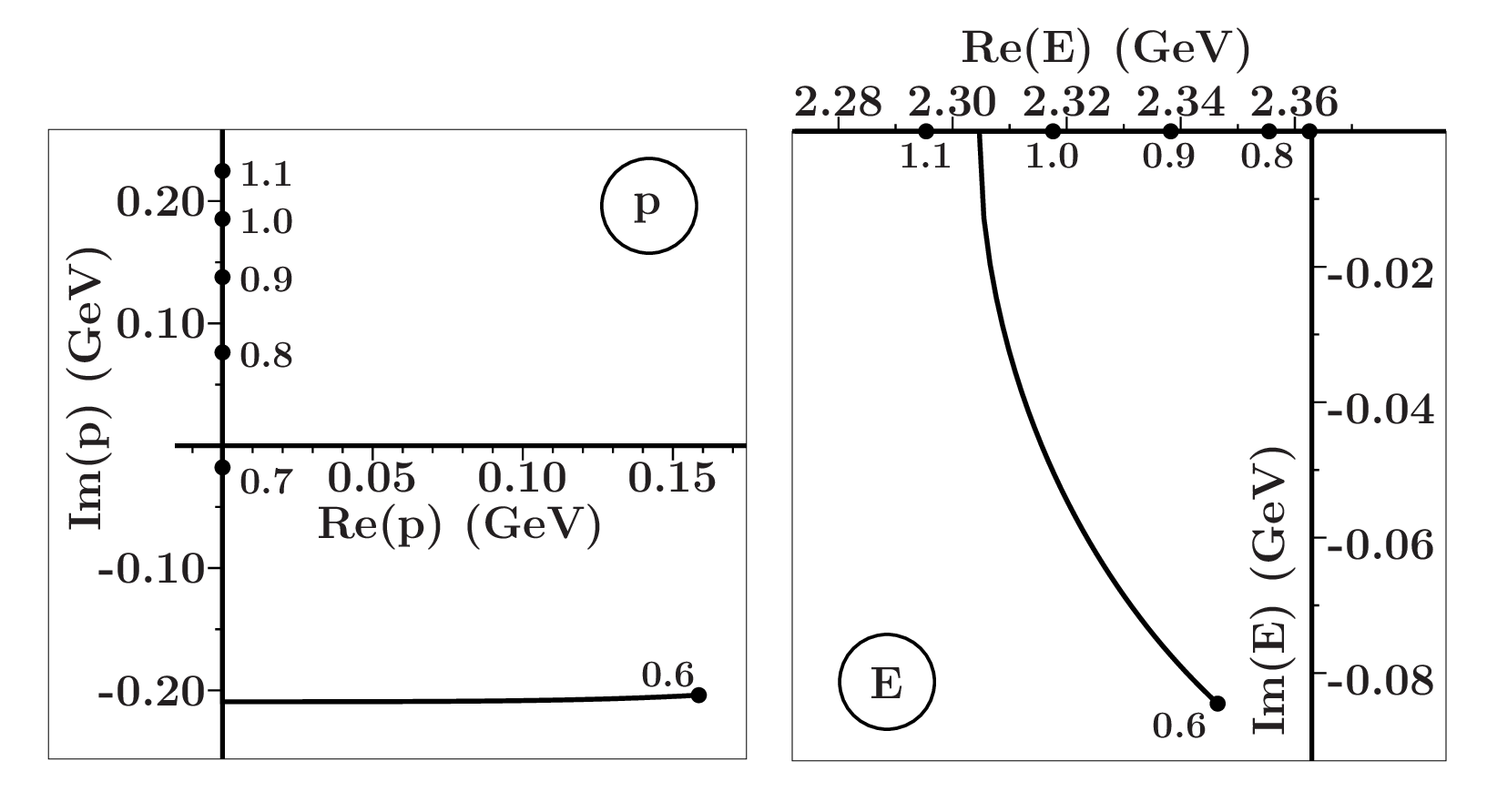}
\caption{Movement of the $D_{s0}^{\ast}$(2317) pole in $S$-wave $DK$ scattering
near the $DK$ threshold, as a function of $\lambda$.
Left: Movement in the $p$ plane.
Right: Movement in the $E$ plane.
We have marked the places where $\lambda$ equals
0.6, 0.7, 0.8, 0.9, 1.0, and 1.1.
On the real axis in the $E$ plane, the pole moves first towards threshold
and then towards smaller real $E$ for increasing $\lambda$.}
\label{threshold}
\end{figure}

{\bf Threshold behavior.}
One may wonder how the lowest poles in Fig.~\ref{KDSpoles}
move along the real axis when $\lambda$ is increased.
This is depicted in Fig.~\ref{threshold}, where
we continuously vary $\lambda$
from a small (0.60) towards a large value (1.2).
In the complex-momentum (p) plane, we observe that the pole
turns purely imaginary somewhere in the lower half-plane.
This is typical for S-wave scattering.
For any higher angular momentum this happens at $p=0$.
The $S$-matrix pole represents a bound-state
when it lies on the positive imaginary axis.
On the negative imaginary axis, it represents a virtual bound state.
In this example, the pole lies on top of threshold for a value of $\lambda$
slightly larger than 0.7.

The pole position in the $E$ plane can be obtained from the pole position
in the $p$ plane through the relation
\begin{displaymath}
E\; =\;\sqrt{p^{2}+m^{2}_{D}}\; +\;\sqrt{p^{2}+m^{2}_{K}}
\;\;\; .
\end{displaymath}
\vspace{10pt}

{\bf Properties.}
The properties of the scattering amplitude following from the
$K$ matrix of Eq.~(\ref{Kmat})
are the following.

\begin{itemize}
\item
Analytic in the total invariant mass.
\item
Proper analytical and smooth behavior at all thresholds.
\item
Unitary for real total invariant mass above and below thresholds.
\item
Symmetric for any complex value of the total invariant mass.
\item
Poles in the scattering matrix may be searched for
in the same fashion as poles in any analytical function.
\item
It contains an infinity
of interfering resonances.
\item
It also allows an arbitrary number (N)
of coupled meson-meson channels.
\item
For small $\lambda$
it exhibits the confinement spectrum $E_{n\ell}$.
\item
For the model's value of $\lambda$, it reproduces the resonances and
bound states of meson-meson scattering.
\item
It contains kinematical Adler-type zeros \cite{HEPPH0412078}
at almost the same invariant masses as the theoretical
Adler zeros \cite{HEPPH0505177}.
\item
It also reproduces automatically the low-lying scalar mesons.
\end{itemize}
\vspace{10pt}

{\bf The parameters.}
The parameters $\lambda$, $a$, and $\omega$ in formula~(\ref{Tij})
represent the probability of quark-pair creation, taken equal for up,
down, strange and zero for the heavy flavors in the present
version of the model, the average distance for quark-pair creation,
and the average level spacing of the $q\bar{q}$ confinement spectrum,
respectively. As expected, $a$ is of the order of half a fermi, and
$\omega$ of the order of 200 MeV. In principle, $\lambda$, $a$, and
$\omega$ should follow from QCD.  Furthermore, in our expression
\begin{equation}
E_{n\ell}\; =\; m_{q}\; +\; m_{\bar{q}}\; +\;
\omega\left( 2n+\ell +\fsml{3}{2}\right)
\;\;\; ,
\label{confspectrum}
\end {equation}
for the confinement spectrum, we must specify the constituent quark
masses. We employ here the values $m_n=0.406$ GeV, $m_s=0.508$~GeV,
$m_c=1.562$ GeV, and $m_b=4.724$ GeV from Ref.~\cite{PRD27p1527}.

Moreover, flavor invariance demands that $\lambda^{2}$ and $a^{2}$
be scaled by the flavor mass, or the reduced quark mass in the
case of different flavors
\cite{allflinv,allscaling}.
Consequently, for quark-pair creation, the average distance
and probability decrease with increasing quark mass, as one naively
would expect.
\vspace{10pt}

{\bf Sum rules for vertex functions.}
The vertex functions, represented by $r_{i}(n)$ in formula (\ref{Tij}),
where $i$ indicates the channel index and $n$ the radial excitation
of the intermediate resonance, contain the full structure of all quantum
numbers, with no parameters involved.
The former can be determined from the recoupling matrix elements
of the harmonic-oscillator wave functions for the effective quarks.
The complete procedure is described in Refs.~\cite{allMMrecouple},
assuming the $^3\!P_0$ mechanism for quark-pair creation.

Some of the resulting vertex functions are shown
in Table~\ref{scalar_recoupling}, for a system with quantum numbers
$J^{PC}=0^{++}$.
\begin{table}[htbp]
\begin{tabular}{||c|c|c|c||}
\hline\hline
\tablehead{1}{c}{b}{meson 1} &
\tablehead{1}{c}{b}{meson 2} &
\tablehead{1}{c}{b}{relative} &
\tablehead{1}{c}{b}{recoupling coefficients}\\
\hline
$(nJLS)_{1}$ $\left( J^{PC}\right)$ & $(nJLS)_{2}$ $\left( J^{PC}\right)$ &
$LS$ & $\{ r(n)\}^{2}\times 4^{n}$\\
\hline\hline & & & \\
$\left( 0,0,0,0\right)$ $\left( 0^{-+}\right)$ &
$\left( 0,0,0,0\right)$ $\left( 0^{-+}\right)$ & $0,0$
& $\frac{1}{24}(n+1)$\\ [10pt]
$\left( 0,0,0,0\right)$ $\left( 0^{-+}\right)$ &
$\left( 1,0,0,0\right)$ $\left( 0^{-+}\right)$ & $0,0$
& $\frac{1}{144}(2n+3)(n-1)^{2}$\\ [10pt]
$\left( 1,0,0,0\right)$ $\left( 0^{-+}\right)$ &
$\left( 1,0,0,0\right)$ $\left( 0^{-+}\right)$ & $0,0$
& $\frac{1}{3456}n(2n+1)(2n+3)(n-3)^{2}$\\ [10pt]
$\left( 0,0,0,0\right)$ $\left( 0^{-+}\right)$ &
$\left( 0,1,1,1\right)$ $\left( 1^{++}\right)$ & $1,1$
& $\frac{1}{6}$\\ [10pt]
$\left( 0,1,0,1\right)$ $\left( 1^{--}\right)$ &
$\left( 0,1,0,1\right)$ $\left( 1^{--}\right)$ & $0,0$
& $\frac{1}{72}(n+1)$\\ [10pt]
$\left( 0,1,0,1\right)$ $\left( 1^{--}\right)$ &
$\left( 0,1,0,1\right)$ $\left( 1^{--}\right)$ & $2,2$
& $\frac{1}{18}(2n+5)$\\ [10pt]
$\left( 0,1,0,1\right)$ $\left( 1^{--}\right)$ &
$\left( 1,1,0,1\right)$ $\left( 1^{--}\right)$ & $0,0$
& $\frac{1}{432}(2n+3)(n-1)^{2}$\\ [10pt]
$\left( 0,1,0,1\right)$ $\left( 1^{--}\right)$ &
$\left( 0,1,2,1\right)$ $\left( 1^{--}\right)$ & $0,0$
& $\frac{1}{540}(2n+3)(2n-5)^{2}$\\ [10pt]
$\left( 0,1,0,1\right)$ $\left( 1^{--}\right)$ &
$\left( 0,1,1,0\right)$ $\left( 1^{+-}\right)$ & $1,1$
& $\frac{1}{6}$\\ [10pt]
$\left( 0,0,1,1\right)$ $\left( 0^{++}\right)$ &
$\left( 0,0,1,1\right)$ $\left( 0^{++}\right)$ & $0,0$
& $\frac{1}{432}(2n+3)(n-3)^{2}$\\ [10pt]
$\left( 0,1,1,1\right)$ $\left( 1^{++}\right)$ &
$\left( 0,1,1,1\right)$ $\left( 1^{++}\right)$ & $0,0$
& $\frac{1}{144}(2n+3)(n-2)^{2}$\\ [10pt]
$\left( 0,1,1,0\right)$ $\left( 1^{+-}\right)$ &
$\left( 0,1,1,0\right)$ $\left( 1^{+-}\right)$ & $0,0$
& $\frac{1}{144}(2n+3)(n-1)^{2}$\\ [10pt]
\hline\hline
\end{tabular}
\caption{Recoupling coefficients as a function of radial excitation $n$,
for scalar ($J=0$, $\ell =1$, $S=1$, $n$) decay into two mesons.
Notice that the recoupling coefficients for $n=0$ add up to one.
This means that, in the harmonic-oscillator approach,
there are no additional two-meson channels which can couple to
the ground state of the confinement spectrum.
For the higher radial excitations, the table is still very incomplete.}
\label{scalar_recoupling}
\end{table}
Each of the two scattering products (mesons) is characterized by its
internal quantum numbers
$n_{\ell}$, $J_{\ell}$, $L_{\ell}$, and $S_{\ell}$ ($\ell =1,2$), i.e.,
the internal radial excitation, total angular momentum (spin of the meson),
orbital angular momentum of the $q\bar{q}$ system, and intrinsic $q\bar{q}$
spin, respectively. The relative motion of the two mesons is characterized
by their relative orbital angular momentum $L$ and total spin $S$,
dictated by the $J^{PC}$ quantum numbers.
The relative radial excitation of the two-meson system
is directly related to $n$ (see Refs.~\cite{allrecouple} for details).

When we determine the recoupling coefficients,
we find that they decrease rapidly for higher radial
excitations $n$ (see Table \ref{scalar_recoupling}),
which implies that the higher terms in the sum over $n$
in formula (\ref{Tij}) are suppressed.
\vspace{10pt}

{\bf \bm{S}-wave scattering.}
In Fig.~\ref{KpiS}c we compare the result
of formula (\ref{Tij}) to the data of Refs.~\cite{NPB133p490,NPB296p493}.
We find a fair agreement for total invariant masses up to 1.6 GeV.
However, we should bear in mind that the LASS data
must have larger error bars for energies above 1.5 GeV than suggested in
Ref.~\cite{NPB296p493}, since most data points fall well outside the Argand
circle. Hence, for higher energies, the model should better not follow the data
too precisely.

\begin{figure}[htbp]
\begin{tabular}{ccc}
\hline
\multicolumn{1}{|c|}{\includegraphics[height=110pt]{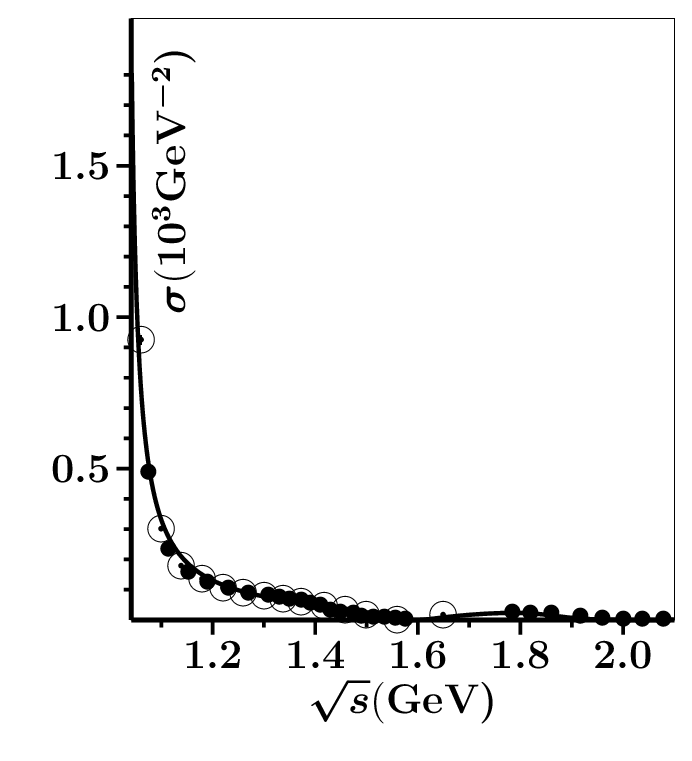}} &
\multicolumn{1}{|c|}{\includegraphics[height=110pt]{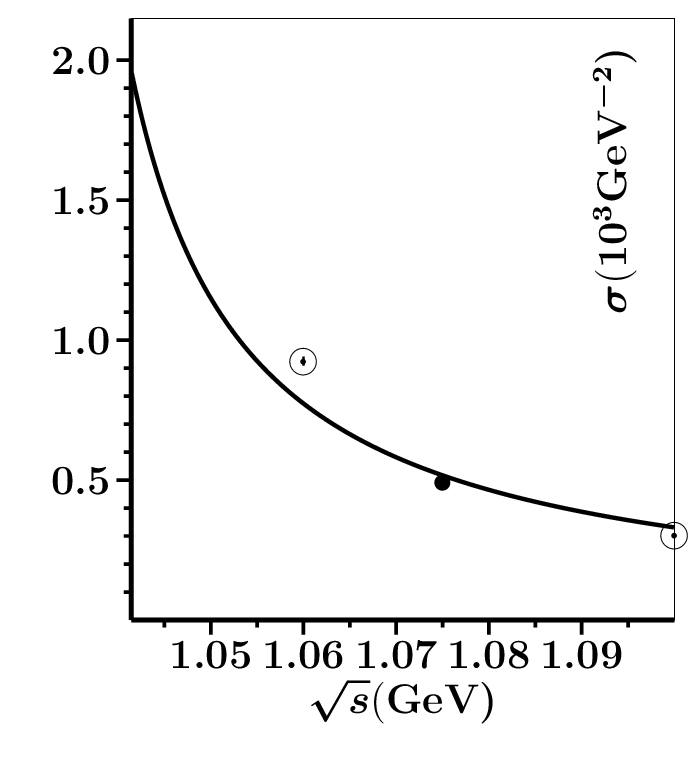}} &
\multicolumn{1}{|c|}{\includegraphics[height=110pt]{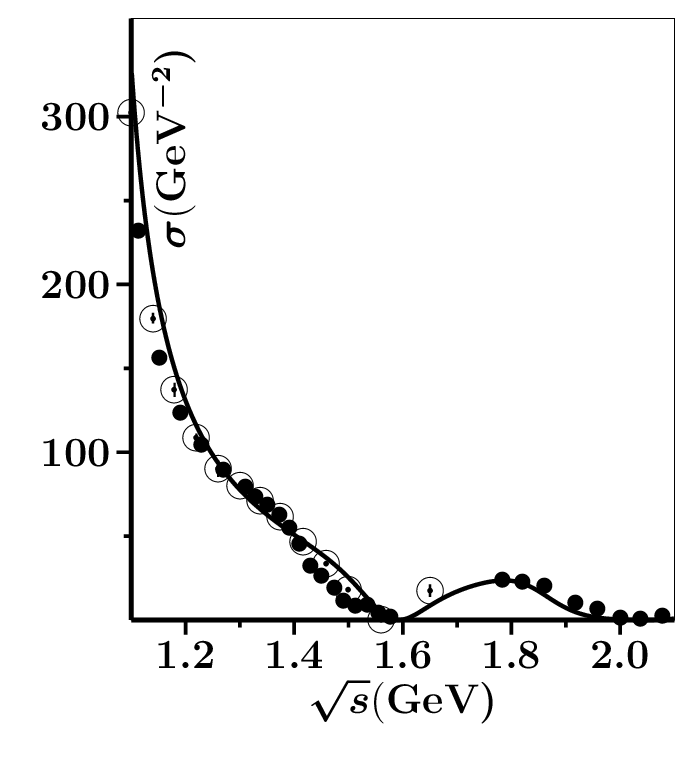}}\\
\hline
(a) & (b) & (c)\\
\end{tabular}
\caption{$S$-wave $K\eta$ ``cross section'' (see text),
as a function of total invariant mass.
({\bf a}): From threshold up to 2.1 GeV.
({\bf b}): Detail for lower energy.
({\bf c}): Detail for higher energy.
The data are taken from Ref.~\cite{NPB133p490} (open circles)
and Ref.~\cite{NPB296p493} (full circles).
}
\label{KetaS}
\end{figure}

\begin{figure}[htbp]
\begin{tabular}{|c|}
\hline
\includegraphics[height=110pt]{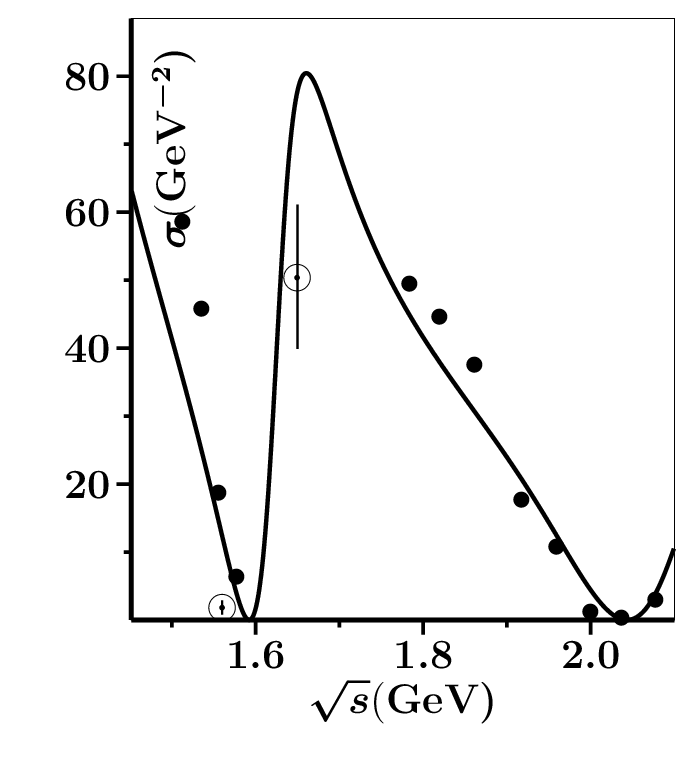}\\
\hline
\end{tabular}
\caption{$S$-wave $K\eta'$ ``cross section'' (see text),
as a function of total invariant mass.
The data are taken from Ref.~\cite{NPB133p490} (open circles)
and Ref.~\cite{NPB296p493} (full circles).
}
\label{KetapS}
\end{figure}

Now, in order to have some idea about the performance of formula
(\ref{Tij}) for $S$-wave $I\!=\!1/2$ $K\pi$ scattering, we argue that,
as in our model there is only one non-trivial eigen-phase shift for the
coupled $K\pi$+$K\eta$+$K\eta'$ system, we may compare the phase shifts
of our model for $K\eta$ and $K\eta'$ to the experimental phase shifts
for $K\pi$.  We do this comparison in Figs.~\ref{KetaS} and \ref{KetapS},
where, instead of the phase shifts, we plot the cross sections,
assuming no inelasticity in either case.
The latter assumption is, of course, a long shot.
Nevertheless, we observe an extremely good agreement.
In particular, for $K\eta'$ (Fig.~\ref{KetapS})
we become aware of a structure in the data at about 1.9 GeV,
indicating the presence of a not-anticipated pole.
This is something we would not have easily noticed from the data alone.

When we inspect formula (\ref{Tij}) for poles
in the $S$-wave isodoublet $K\pi$ scattering amplitude,
then we find the pole structure as summarized in Table~\ref{KpiSpoles},
i.e., five poles at energies up to about 2.2 GeV real part.
The first pole, at $0.772 - 0.281 i$ GeV, describes the heavily disputed
$K^{\ast}_{0}(800)$ structure,
whereas the second pole, at $1.52 - 0.097 i$ GeV, represents the
well-established $K^{\ast}_{0}(1430)$ resonance.
\begin{table}[htbp]
\begin{tabular}{||c|c|c|c|c|c||}
\hline\hline
Pole (GeV) & $0.772 - 0.281 i$ & $1.52 - 0.097 i$ &
$1.79 - 0.052 i$ & $2.04 - 0.15 i$ & $2.14 - 0.065 i$\\
Origin & continuum & confinement & confinement & continuum & confinement\\
\hline\hline
\end{tabular}
\caption{$T$-matrix poles for $S$-wave $K\pi$ scattering,
as obtained from Eq.~(\ref{Tij}).}
\label{KpiSpoles}
\end{table}

Our model is explicitly flavor-independent, meaning that the only flavor
breaking in formula~(\ref{Tij}) stems from the effective quark masses
which determine the ground state of the confinement spectrum
(see Eq.~\ref{confspectrum}), and from the masses of the mesons in the
scattering channels. Consequently, $\pi\pi$ scattering is not very
different from $K\pi$ scattering in our model. We may expect then that
each of the two flavor combinations that couple to isoscalar $S$-wave
$\pi\pi$ and $KK$ scattering has a pole structure similar to the one
in isodoublet $K\pi$ scattering, with the proviso that
$n\bar{n}$-$s\bar{s}$ mixing in the $I\!=\!0$ case introduces an extra
complication.
\begin{table}[htbp]
\begin{tabular}{||c|c|c|c|c|c||}
\hline\hline
$n\bar{n}$ (GeV) & 0.66 & 1.42 & 1.69 & 1.94 & 2.04\\
$s\bar{s}$ (GeV) & 0.86 & 1.62 & 1.89 & 2.14 & 2.24\\
\hline\hline
\end{tabular}
$\;\;\;$
\begin{tabular}{||c|c|c|c|c|c||}
\hline\hline
$f_{0}$ (GeV) & 0.6 & 1.37 & 1.71 & 2.02 & 2.33\\
$f_{0}$ (GeV) & 0.98 & 1.50 & & 2.20 & \\
\hline\hline
\end{tabular}
\caption{
The left-hand table shows
the real parts of the poles in the amplitude
for $n\bar{n}$ (upper row) $s\bar{s}$ (lower row)
$S$-wave $\pi\pi$+$KK$ coupled-channel scattering,
as expected in analogy with $K\pi$ (see Table~\ref{KpiSpoles}),
whereas the right-hand table shows
the observed $f_{0}$ states listed in
the Tables of Particle Properties \cite{PLB592p1}.
In the upper (lower) rows, we collect those resonances which we
expect to be mainly $n\bar{n}$ ($s\bar{s}$).
}
\label{isoscalar}
\end{table}
The to-be-expected poles are given in Table~\ref{isoscalar},
alongside the $f_{0}$ structures reported in experiment.

The often read comment that {\it too many isoscalar states are observed}
\/\cite{PLB541p22}, in order to justify the application of alternative quark,
or even quarkless, configurations \cite{HEPPH0411396}, is not confirmed here.

Most probably, mesons are just mixtures \cite{PLB600p223} of
{\it quark-antiquark} \/states \cite{allqqbar},
{\it two-meson molecules} \/\cite{allmolecules},
{\it glueballs} \/\cite{allglueball},
{\it tetraquarks} \/\cite{alltetra},
{\it hexaquarks},
{\it hybrids} \/\cite{HEPPH0211289},
and so forth.
Here, we have shown that the first two of the latter list of possible
components are the most relevant ones. Moreover, a resonance is really
a collection of states, all with different masses. Each of these states
will have a different composition.

In Table~\ref{Swave} we summarize the scalar poles which we obtained
in the past \cite{ZPC30page615,allscalar,allflinv}. The first four lines
in the third column of Table~\ref{Swave} are refered to in the literature
as the scalar-meson nonet \cite{allnonet},
given by the isoscalars $f_{0}(600)$ and $f_{0}(980)$,
the isotriplet $a_{0}(980)$, and the isodoublet pair $K_{0}^{\ast}(800)$
\cite{allkappa}. Here, they form part of the complex pole spectrum of
the general scattering amplitude (\ref{Kmat}), where they appear as the
lowest-lying continuum poles.
The first four lines in the fourth column of Table~\ref{Swave}
describe the nonet of scalar mesons that stem directly
from the $J^{P}=0^{+}$ ground states of the confinement spectrum.
\begin{table}[htbp]
\begin{tabular}{||c|c|c|c|c||}
\hline\hline
\tablehead{1}{c}{b}{channel} &
\tablehead{1}{c}{b}{\bm{q}-\bm{\bar{q}}} &
\tablehead{1}{c}{b}{continuum} &
\tablehead{1}{c}{b}{ground state} &
\tablehead{1}{c}{b}{excitation}\\
\hline &  &
\multicolumn{1}{r|}{GeV} &
\multicolumn{1}{r|}{GeV} &
\multicolumn{1}{r|}{GeV}\\
$\pi\pi$ & $n$-$\bar{n}$ & $0.47 - 0.21 i$ & $1.36 - 0.13 i$ & -\\
$K\pi$ & $n$-$\bar{s}$ & $0.77 - 0.28 i$ & $1.52 - 0.10 i$ & $1.79 - 0.05 i$\\
$\eta\pi$  & $n$-$\bar{n}$ & $0.97 - 0.028 i$ & $1.45 - 0.13 i$ & -\\
$\pi\pi$ & $s$-$\bar{s}$ & $0.99 - 0.020 i$ & $1.51 - 0.06 i$ & -\\
$D\pi$ & $c$-$\bar{n}$ & $2.14 - 0.16 i$ & $2.58 - 0.12 i$ & -\\
$D K$ & $c$-$\bar{s}$ & $2.33$ & $2.80 - 0.20 i$ & -\\
$B\pi$ & $n$-$\bar{b}$ & $6.06 - 0.29 i$ & $5.46 - 0.03 i$ & $6.03 - 0.05 i$\\
$B K$ & $s$-$\bar{b}$ & $6.21 - 0.33 i$ & $5.61$ & $6.05 - 0.03 i$\\
$B D$ & $c$-$\bar{b}$ & $7.12 - 0.43 i$ & $6.64$ & $7.11 - 0.03 i$\\
\hline\hline
\end{tabular}
\caption{$S$-wave scattering poles for various flavor/isospin
combinations with $J^{P}=0^{+}$.
In the columns ``continuum'', ``ground state'', and ``excitation'' we indicate
the origin of the poles, i.e., either meson-meson continuum,
or confinement spectrum (ground state or first radial excitation).}
\label{Swave}
\end{table}
\vspace{10pt}

{\bf Summary and discussion.}
In the foregoing, we have shown that the most relevant information on
(meson) spectra amounts to the knowledge about
the complex pole positions of the scattering amplitudes.
Furthermore, we presented a general form of amplitudes for non-exotic
hadron-hadron scattering with any possible flavor combination,
based upon the most prominent properties of strong interactions,
namely {\it quark confinement}, {\it quark-pair creation},
and {\it flavor invariance}.
The resulting pole spectrum may very well be compared to experiment.
It is true that the predicted cross sections and phase shifts
are not always in very accurate agreement with the data, as could hardly be
expected in view of the extremely wide scope of the model. In part this may
also be due to the inaccuracy of some experimental numbers.
Nevertheless, the model can be improved as well, as we suggest next.
\begin{itemize}
\item
At present, we determine the vertices in the framework of the $^3\!P_0$
mechanism, assuming that all quark masses involved are the same.
However, different quark masses can be dealt with, too.
Computer programs already exist for several years,
but have not yet been implemented.
\item
The one-delta-shell approximation ``unfortunately'' works too well.
Hence, although it is perfectly known how to implement
more complicated transition potentials \cite{CPC27p377},
and even some computer code is ready,
the finishing touch could still take a while,
unless some extra manpower becomes available.
\item
The kinematics of meson pairs below threshold is most certainly not
being dealt with in the most appropriate way by us. This is a subject
which deserves to be studied in more detail.
\end{itemize}
\vspace{10pt}

{\bf Acknowledgment.}
One of us (EvB) wishes to thank the organizers
for inviting him to the Hadron05 conference,
their warm hospitality, and
for organizing this great event.
This work was partly supported by the
{\it Funda\c{c}\~{a}o para a Ci\^{e}ncia e a Tecnologia}
of the {\it Minist\'{e}rio da Ci\^{e}ncia, Tecnologia e Ensino Superior}
\/of Portugal, under contract POCTI/\-FP/\-FNU/\-50328/\-2003 and grant
SFRH/BPD/9480/2002.

\end{document}